\documentclass[pop,superscriptaddress,showpacs,amsmath,amssymb,reprint]{revtex4-1}
\usepackage{graphicx,dcolumn,bm}

\newlength{\plotwidth}
\newcommand{\e}{\eqref}

\begin{document}


\title{Vlasov multi-dimensional model dispersion relation}

\author{Pavel M. Lushnikov}
\email{plushnik@math.unm.edu}
\affiliation{Department on Mathematics and Statistics, University of New Mexico, New Mexico 87131, USA}
\author{Harey A. Rose}
\affiliation{New Mexico Consortium, Los Alamos, New Mexico 87544, USA}
\author{Denis A. Silantyev}
\affiliation{Department on Mathematics and Statistics, University of New Mexico, New Mexico 87131, USA}
\affiliation{New Mexico Consortium, Los Alamos, New Mexico 87544, USA}
\author{Natalia Vladimirova}
\affiliation{Department on Mathematics and Statistics, University of New Mexico, New Mexico 87131, USA}
\affiliation{New Mexico Consortium, Los Alamos, New Mexico 87544, USA}

\begin{abstract}
A hybrid model of the Vlasov equation in multiple spatial dimension $D>1$ [H. A. Rose and W. Daughton, Physics of Plasmas \textbf{18}, 122109 (2011)], the Vlasov multi dimensional model (VMD), consists of standard Vlasov dynamics along a preferred direction, the $z$ direction, and $N$ flows. At each $z$ these flows are in the plane perpendicular to the $z$ axis. They satisfy Eulerian-type hydrodynamics with coupling by self-consistent electric and magnetic fields. Every solution of the VMD is an exact solution of the original Vlasov equation.
We show convergence of the VMD Langmuir wave dispersion relation in thermal plasma to that of Vlasov-Landau as $N$ increases. Rotational symmetry about the $z$ axis in $3D$ of small
perpendicular wavenumber Langmuir fluctuations is demonstrated for $N\geq 6$, with flows arranged uniformly over the azimuthal angle.
\end{abstract}

\pacs{52.65.-y, 52.65.Ff, 52.38.-r, 52.35.-g}

\maketitle

\section{Introduction}

Multi-dimensional simulations of the Vlasov equation \cite{Vlasov} are possible \cite{Winjum}, but present a significant computational burden compared to particle in cell (PIC) methods \cite{Dawson}. Modeling the Vlasov equation as a fluid is inadequate for linear regimes in which Landau damping \cite{Landau} and, nonlinearly, e.g., electron trapping
by Langmuir waves (LWs), are significant. However, if plasma wave propagation is largely confined to a narrow cone about  $z$ axis, as may be the case for stimulated Raman scattering (SRS)
\cite{DuBois,Mont,Russell}, then these kinetic effects may be adequately described by a recently developed kinetic-fluid hybrid  Vlasov multi dimensional model (VMD) \cite{Rose},
consisting of standard Vlasov dynamics along the $z$ direction and $N$ fluid flows in the $xy$ (perpendicular) plane, coupled by self-consistent electric and magnetic fields.
Each flow convects its corresponding one-dimensional ($1D$) distribution function,$\left\lbrace f_i(x,y,z,v_z,t)\right\rbrace_{i=1\ldots N} $, in the $xy$ plane.

In this paper we show limits of
convergence and stability of the VMD dispersion relation, in thermal plasma, to that of the Vlasov equation
dispersion relation including convergence of LW branch as $N$ increases. We consider both a two-dimensional ($2D$) case with only one
transverse direction $x$ taken into account and a full three-dimensional $(3D)$ case with two transverse directions
$x$ and $y$. Rotational invariance about the $z$ axis in $3D$ for small transverse wavenumber fluctuations is recovered for $N\geq 6$, with flows arranged uniformly over the azimuthal angle.

The paper is organized as follows. In Section \ref{sec:VMDMODELREDIX} we recall the basic properties of VMD. In Section \ref{sec:ACCURACYWITHFLOWNUMBERIN2D} we analyze the
increase of VMD accuracy with the increase of the number of flows $N$ as well as the boundaries of stability.
In Section \ref{sec:VMDDISPERSIONRELATIONIN3D} the properties of $3D$ VMD dispersion relation are studied.
In Section \ref{sec:Conclusion} the main results of the paper are
discussed.

\section{VMD MODEL REDIX}
\label{sec:VMDMODELREDIX}

In this Section, the VMD  basic properties \cite{Rose} are recalled. Let $g$ be a particular species' phase space distribution function. The VMD ansatz is:
\begin{equation}
    g({\bf x},{\bf v},t) =\sum_{i=1}^N{f_i({\bf x},v_z,t)\delta({\bf v}_{\bot}-{\bf u}_{i\bot}({\bf x},t)  )}.
    \label{eq:distr}
\end{equation}
Here $3D$ position and velocity vectors are denoted by {\bf x} and {\bf v}, respectively, and the latter may represented by its perpendicular ($xy$ plane), ${\bf v}_{\bot}$, and parallel ($z$ axis) projections, $v_z$,
\begin{equation}
    {\bf v}={\bf v}_{\bot}+v_z\hat{{\bf e}}_z, \quad  {\bf v}_{\bot}\cdot\hat{{\bf e}}_z=0,
    \label{eq:decomposition}
\end{equation}
with $\hat{{\bf e}}_z$ being the unit vector along the $z$ axis. The Vlasov equation, in units such that electron mass and charge are normalized to unity, is
\begin{equation}
    \left\lbrace\frac{\partial }{\partial t} + {\bf v}\cdot\nabla+ {\bf E}\cdot \frac{\partial }{\partial {\bf v}} \right\rbrace g=0,
    \label{eq:vlasov}
\end{equation}
where $\bf E$ is the electric field. Magnetic field effects are ignored for clarity. It can be shown  \cite{Rose} that Eqs. (\ref{eq:distr}) and (\ref{eq:vlasov}) imply
\begin{equation}
    \left\lbrace\frac{\partial }{\partial t} + v_z\frac{\partial }{\partial z}+ E_z\frac{\partial }{\partial v_z} \right\rbrace f_i+\nabla_\bot\cdot({\bf u}_{i\bot}f_i)=0,
    \label{eq:vmd1}
\end{equation}
\begin{equation}
    \left\lbrace\frac{\partial }{\partial t} + {\bf u}_{i}\cdot\nabla  \right\rbrace {\bf u}_{i\bot} ={\bf E}_\bot.
    \label{eq:vmd2}
\end{equation}
Flow fields ${\bf u}_i$ are decomposed similar to \eqref{eq:decomposition} as
\begin{equation}
    {\bf u}_i={\bf u}_{i\bot}+u_{iz}\hat{{\bf e}}_z, \quad  {\bf u}_{i\bot}\cdot\hat{{\bf e}}_z=0, \quad i=1\ldots N
    \label{eq:udecomposition}
\end{equation}
with their perpendicular components determined by Eq.(\ref{eq:vmd2}) and its $z$ component by
\begin{equation}
    \rho_i u_{iz}=\int v_z f_i({\bf x},v_z,t)  dv_z,
    \label{eq:u_zi}
\end{equation}
where
 \begin{equation}
    \rho_i =\int  f_i({\bf x},v_z,t)  dv_z
    \label{eq:density}
\end{equation}
is the density of $i$th flow.

The electric field ${\bf E}$ is determined by a total density $\rho=\sum_{i=1}^N\rho_i$ together with the Maxwell's equations. Recall that the summation over $i$ is a sum over flow field components for a given species.  An additional sum over species is required to obtain the total charge and current densities but below we assume a single specie plasma for simplicity. Eqs.(\ref{eq:vmd1}) and (\ref{eq:vmd2}) constitute the VMD model. VMD solutions, for wave propagation strictly along the $z$ axis, are precisely correct because in this case the VMD model coincides with Vlasov equation.

Ignoring magnetic field, the Maxwell's equations are reduced in   the electrostatic regime to
\begin{align} \label{phidef}
{\bf E}=-\nabla \phi,
\end{align}
 with the Poisson equation
\begin{align} \label{Poisson}
\nabla^2\phi=-\rho,
\end{align}
where $\phi$  is the electrostatic potential and the factor $4\pi$ is absent in \eqref{Poisson}
because we normalized the
length to the electron Debye length, and frequency to the
electron plasma frequency \cite{Rose}.

In Ref. \cite{Rose} it was shown that the VMD model's basic limitation in $2D$ with $N=2$ is the restriction to wave propagation dominantly along the $z$ axis to avoid an unphysical two-stream-instability. These results are recalled and extended to large values of $N$ in the next Section.

\section{INCREASE OF ACCURACY WITH FLOW NUMBER IN 2D}
\label{sec:ACCURACYWITHFLOWNUMBERIN2D}
In 2D, for the minimal $N=2$ case, linearizing equations (\ref{eq:vmd1}),(\ref{eq:vmd2}), \eqref{phidef} and \eqref{Poisson} around a ``thermal equilibrium'' distribution function,
\begin{equation}
\begin{split}
    g_0=\frac{1}{2}[\delta(v_x-u)+\delta(v_x+u)]f_0(v_z),  \\
     \qquad f_0(v)=\frac{1}{\sqrt{2\pi}}\exp\left (-\frac{v^2}{2}\right ),
\end{split}
    \label{eq:distr2d}
\end{equation}
and assuming that perturbation $\propto \exp{(i{\bf k}\cdot {\bf x}-i\omega t)}$ one obtains VMD  dispersion relation for a fluctuation with wavenumber $k=|{\bf k}|$, making an angle $\theta$ with respect to the $z$ axis
\begin{equation}
\begin{split}
    &4k^2=Z'(\zeta_+)+Z'(\zeta_-)\\
    &\qquad \qquad-\tan^2(\theta)\left(\frac{Z(\zeta_+)}{\zeta_+} + \frac{Z(\zeta_-)}{\zeta_-} \right),   \\
    &\zeta_\pm=\frac{\mp u \sin(\theta) + \omega/k}{\sqrt{2}\cos(\theta)}, \\    &Z(\zeta)= \frac{1}{\sqrt{\pi}} \int^{+\infty}_{-\infty}  \frac{\exp(-t^2)}{t-\zeta} dt,
\end{split}\label{eq:vmd2Disp}
\end{equation}
where $Z$ is the plasma dispersion function \cite{Fried}.

In comparison, a linearization of the Vlasov equation \eqref{eq:vlasov} with
\eqref{phidef} and \eqref{Poisson} around the  ``thermal equilibrium'' \eqref{eq:distr2d} results in the following dispersion relation

\begin{equation}
\begin{split}
    &4k^2\cos^2(\theta)=Z'(\zeta_+)+Z'(\zeta_-),
\end{split}\label{eq:vlasov2Disp}
\end{equation}
where $\zeta_\pm$ are defined in  \eqref{eq:distr2d}.

 VMD dispersion relation  \eqref{eq:vmd2Disp} coincides with the Vlasov equation dispersion relation \eqref{eq:vlasov2Disp}  in two limiting cases when $\theta\rightarrow0$ and $\theta\rightarrow\pi/2$.
 In the case of $\theta\rightarrow0$ VMD dispersion relation coincide with the Vlasov dispersion relation for isotropic thermal plasma:
\begin{align} \label{isotropicVlasovdispersion}
2k^2=Z'(v_\varphi/\sqrt{2}),
\end{align}
where $v_\varphi\equiv\omega/k$ is the phase velocity. And in the second case of $\theta\rightarrow\pi/2$  we obtain a well-known cold plasma two-stream dispersion relation \cite{NicholsonBook1983}
\begin{equation}
    2=\frac{1}{(\omega-uk)^2} + \frac{1}{(\omega+uk)^2}.
    \label{eq:twostream}
\end{equation}
It has been shown \cite{Rose} that LW dispersion relation \eqref{isotropicVlasovdispersion} is qualitatively recovered in VMD dispersion relation \eqref{eq:vmd2Disp} for $u\approx1$ and angle of propagation between the $z$ axis and wave-vector ${\bf k}$, $\theta\lesssim0.65$, beyond which a variant of the two-stream instability is encountered. This region of instability is an artifact of the two-stream model, Eq. (\ref{eq:distr2d}). Here we demonstrate that an increase of $N$ in $2D$ has a stabilizing effect.

To emulate thermal equilibrium, with $N$ transverse flows, $\{u_i\}_{i=1\ldots N}$, it is natural to choose their weights, $\{\rho_i\}_{i=1\ldots N}$, proportional to that of true thermal equilibrium,
\begin{equation}
\begin{split}
     g_0(v_x,v_z)=f_0(v_z) \sum_{i=1}^N \rho_i \delta(v_x-u_i), \\
  \sum_{i=1}^N\rho_i =1,\quad \rho_i \sim \exp(-\frac{u_i^2}{2}),
   \label{eq:distr2dN}
\end{split}
\end{equation}
where the total density $\rho$  is  normalized to unity. The apparent factorization implies independence of $v_z$ and $v_x$ fluctuations, as in true thermal equilibrium. Aside from some general properties such as symmetry about $v_x=0$, it is not clear what choice of flows optimally recovers properties of exact Vlasov solutions. One additional restriction that might be imposed is having unit mean square transverse velocity,
\begin{equation}
    \sum_{i=1}^N \rho_i u_i^2=1,
    \label{eq:meansquareone}
\end{equation}
so that the transverse and $z$-direction temperatures coincide.
Criteria for choosing $u$ for $N=2$, Eq. (\ref{eq:distr2d}), were presented in Ref. \cite{Rose}. Here we emphasize $N\gg 1$ results. The general VMD dispersion relation for $2D$ case with $N\geq 2$ is
\begin{equation}
\begin{split}
    2k^2=\sum_{i=1}^N \rho_i \left( Z'(\zeta_i)-\tan^2(\theta) \frac{Z(\zeta_i)}{\zeta_i} \right), \\
     \zeta_i=\frac{-u_i \sin(\theta) + \omega/k}{\sqrt{2}\cos(\theta)}.
\end{split}
    \label{eq:vmd2DNDisp}
\end{equation}

\begin{figure}
\includegraphics[width=3.4in]{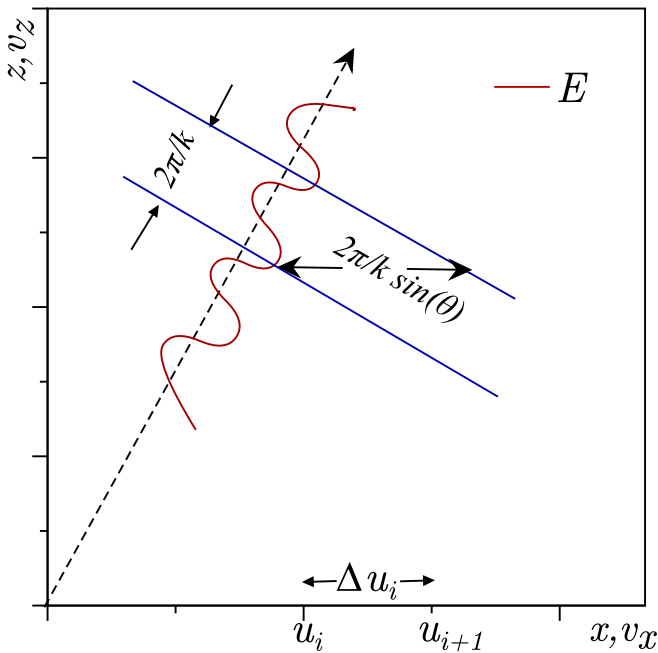}
\caption{Cartoon of electric field, red oscillatory curve, and phase fronts, blue lines, illustrates possible electron de-phasing. A VMD electron typically has  a velocity difference in $v_x$ direction  with the wave's, $v_\varphi\sin(\theta)$, of order
$\Delta u_i$ and will dephase in a time of order $2\pi/[k\Delta u_i\sin(\theta)$].}
\label{fig:dephasing}
\end{figure}

One reasonable flow requirement is a recovery of standard Landau damping in the linear regime and the other one is to ensure a maximum region of stability. Recall the test-particle point of view of Dawson \cite{Dawson2} for Landau damping where an electron with a
LW's phase velocity $v_\varphi$ interacts with the wave for a finite time. However, for the VMD  to recover a LW's Landau damping, $\nu_L$, it is sufficient (but not necessary, see Fig. 3 in
Ref. \cite{Rose}) that its phase velocity is close to a kinematically accessible one, as in Fig. \ref{fig:dephasing} which illustrates a plane wave sinusoidal electric field,
with wavenumber $k$, phase velocity $v_\varphi$, propagating at an angle $\theta$ with the $z$ axis, and whose $x$ velocity projection, $v_{\varphi x}=v_\varphi\sin(\theta)$, lies in
the interval $(u_i,u_{i+1})$. Typically, $v_\varphi$ will not coincide with a particular $u_i$, rather, a test particle horizontal velocity will differ from precise resonance
with the wave by order $\Delta u_i\equiv u_{i+1}-u_{i}$ causing its location to dephase with the wave in a time scale $\sim1/[k\Delta u_i\sin(\theta)]$.
To obtain a qualitatively faithful rendition of Landau damping this must be large compared to a Landau damping time, or equivalently
\begin{equation}
    k \Delta u_i\sin(\theta)\ll2\pi\nu_L.
    \label{eq:dephcond}
\end{equation}

\begin{figure}
\includegraphics[width=3.5in]{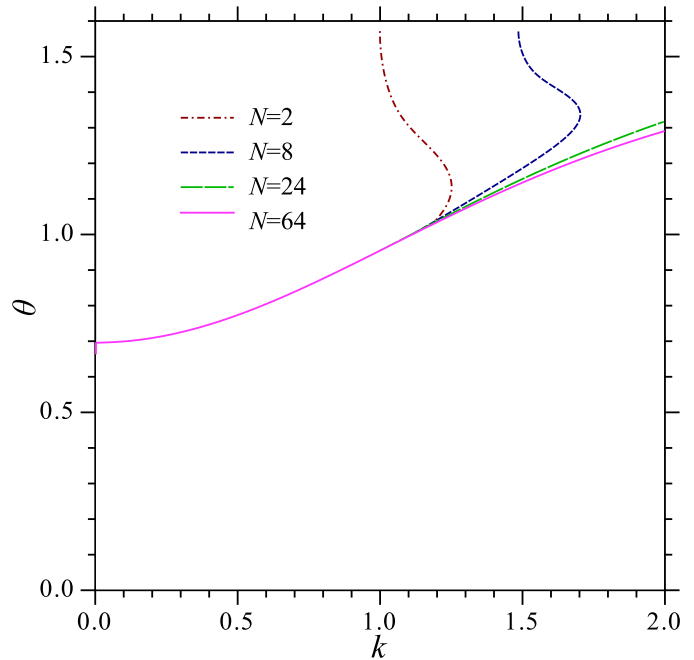}
\caption{The stability boundary for VMD dispersion relation \e{eq:vmd2DNDisp}
for different  number of flows, $N$. Stable region is below and to the right of the corresponding curve.
As  $N$ increases, the stability boundary converges to the universal curve. Above this boundary, instability persists, but growth rate goes to zero as $\sim1/N^{0.5}$.}
\label{fig:boundary2DN}
\end{figure}

There is a big freedom in different choices of flow spacing.
We  focus on a particular family of flows but we also tested other families and found similar results.
For each member of that family we consider  VMD dispersion relation \e{eq:vmd2DNDisp}   in two ways: (a) we find range of parameters when \e{eq:vmd2DNDisp} gives
only stable solutions and (b) we compare solutions of  LW branch of VMD dispersion relation \e{eq:vmd2DNDisp} with the exact result for
 the LW branch of the Vlasov equation dispersion relation \e{isotropicVlasovdispersion}.

We choose a flow spacing, $\Delta u=|u_{i+1}-u_i|$ which is independent of flow index, $i$.
 We decrease $\Delta u$ with the increase of $N$ as follows. Consider $N$ flows $u_i$ evenly spaced between
 $-U_{max}$ and $U_{max}$, skipping $u=0$.  Take a positive integer $n$ and choose
\begin{equation}
    U_{max}=n,  \qquad \Delta u=1/2^{n-1} ,\qquad  N=n2^n.
    \label{eq:u}
\end{equation}
According to \eqref{eq:u}, with the increase of $n$, $U_{max}$ grows linearly,  $\Delta u$ decreases exponentially and $N$ grows  faster by a factor $n$ than the exponential growth $2^n$.

(a) The stability boundary of VMD dispersion relation \e{eq:vmd2DNDisp} is shown in Fig. \ref{fig:boundary2DN} for $N=2,8,24$, and $64$ ($n=1,2,3,4$).
For each given $N$ the region below the corresponding curve has only stable modes $Im(\omega)\le 0$ and above the curve there is at least one unstable solution
$Im(\omega)> 0$. While the area of the unstable region  increases (but converges) with increase of $N$, the maximum growth rate $\max{Im(\omega)}$ decreases as $1/N^\alpha$,
with $\alpha\approx0.5$  for the above scheme  \eqref{eq:u}.
For this family of flows unstable roots of \e{eq:vmd2DNDisp} for $\theta\neq0$ correspond to ${Re}(\omega)\approx0$. These unstable roots move as $k$ and $\theta$ are changed.
Stability boundary is obtained by finding numerically $k$ such that for each given $\theta$ the most unstable solution of the dispersion relation turns into $\text{Im}(\omega)=0$.
The solutions of \eqref{eq:vmd2DNDisp} corresponding to Langmuir wave are stable for any choice of $k$ and $\theta$.

\begin{figure}
\includegraphics[width=\plotwidth]{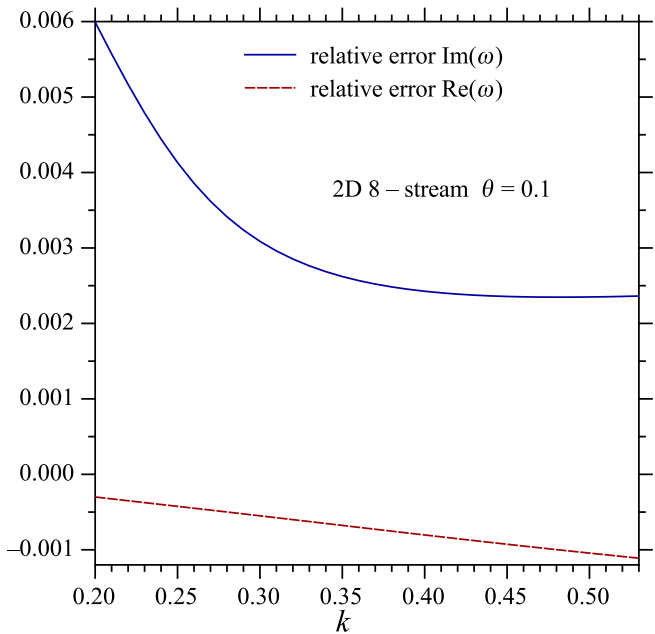}
\caption{Relative error between LW branch of VMD dispersion relation \e{eq:vmd2DNDisp} and  LW dispersion relation  \e{isotropicVlasovdispersion} for $N=8$ and  $\theta=0.1$ as a function $k$.}
\label{fig:subfigure1}
\end{figure}

\begin{figure}
\includegraphics[width=\plotwidth]{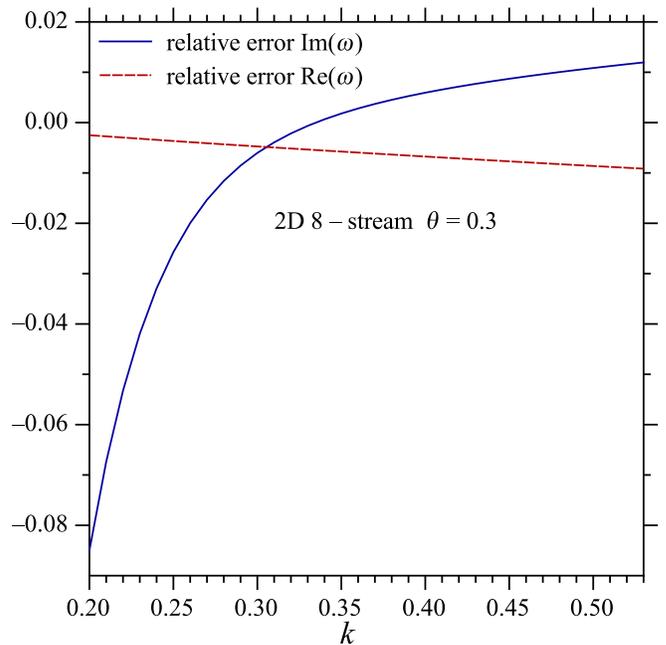}
\caption{Relative error between LW branch of VMD dispersion relation \e{eq:vmd2DNDisp} and  LW dispersion relation  \e{isotropicVlasovdispersion} for $N=8$ and  $\theta=0.3$.}
\label{fig:subfigure2}
\end{figure}

\begin{figure}
\includegraphics[width=3.3in]{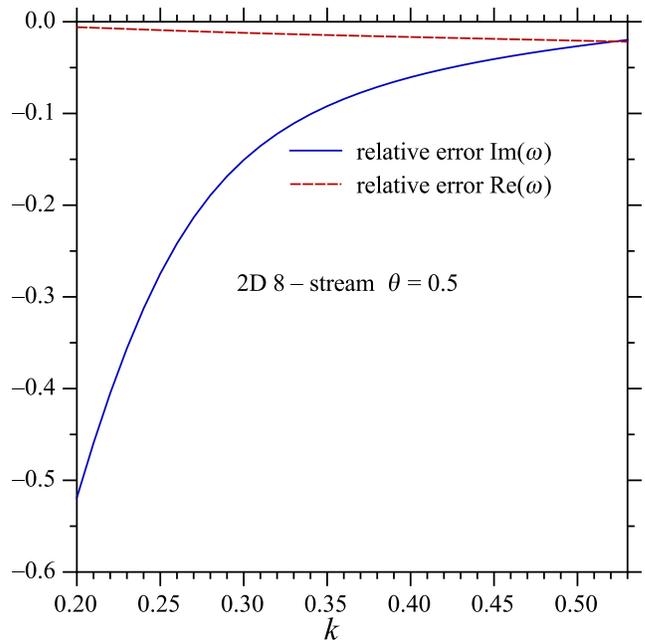}
\caption{Relative error between LW branch of VMD dispersion relation \e{eq:vmd2DNDisp} and  LW dispersion relation  \e{isotropicVlasovdispersion} for $N=8$ and  $\theta=0.5$.}
\label{fig:subfigure3}
\end{figure}

\begin{figure}
\includegraphics[width=3.3in]{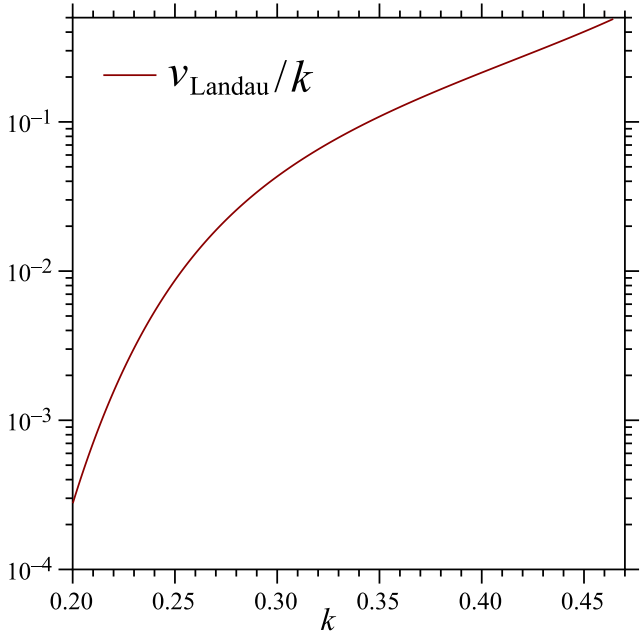}
\caption{A dependence of   $\nu_L/k$ on $k$ in LW dispersion relation  \e{isotropicVlasovdispersion}.
}
\label{fig:subfigure4}
\end{figure}

(b)  Roots of LW branch of VMD  dispersion relation  \eqref{eq:vmd2DNDisp}  for $N=8$, $u_i=[-2,-1.5,-1,-0.5,0.5,1,1.5,2]$, are consistent with the estimate
 (\ref{eq:dephcond}). Figs. \ref{fig:subfigure1}, \ref{fig:subfigure2} and \ref{fig:subfigure3} show the relative error  between LW branch of VMD dispersion relation \e{eq:vmd2DNDisp} and
 the exact LW dispersion relation  \e{isotropicVlasovdispersion} for $\theta=0.1, \, 0.3$ and $0.5$ respectively. LW branch of VMD  dispersion relation for $\theta\neq0$ is obtained by
 continuation of LW branch  \e{isotropicVlasovdispersion}   identified at $\theta=0$. Note that the error in $\text{Im}(\omega)$ (blue solid curves) grows rapidly as $k$ decreases below $0.3$,
 consistent with the rapid decrease of $\nu_L/k$ with decrease of $k$ as shown in Fig. \ref{fig:subfigure4}.
 That rapid decrease provides strong constraints on flow spacing, $\Delta u$, and angle of propagation, $\theta$, with respect to the $z$ axis, as per Eq. (\ref{eq:dephcond}).

\section{VMD DISPERSION RELATION IN $3D$ }
\label{sec:VMDDISPERSIONRELATIONIN3D}
\begin{figure}
          \includegraphics[width=\plotwidth]{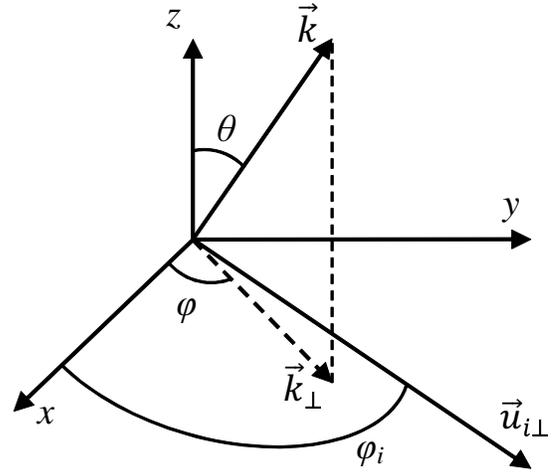}
          \caption{Schematic $3D$ VMD geometry, with only one of the equilibrium flows, ${\bf u}_{i\perp}$,  depicted.}
          \label{fig:3Dvectors}

\end{figure}

Consider VMD thermal equilibrium distribution function of the form (\ref{eq:distr2d}) generalized to $3D$ with $N$ equally weighted flows in transverse plane arranged uniformly over the azimuthal angle:
\begin{equation}
    g_0({\bf v}_\perp,v_z)=\frac{1}{N}f_0(v_z)\sum_{i=1}^N{\delta({\bf v}_{\bot}-{\bf u}_{i\bot})}
    \label{eq:distr3d}.
\end{equation}
The magnitude of ${\bf u}_{i\bot}$ for each flow is $\sqrt{2}$, so that $\langle({\bf v}\cdot{\bf n})^2\rangle=1$ for any N$\geq$3 independent of the unit vector, ${\bf n}$, direction with angular brackets indicating average over the VMD thermal equilibrium state (\ref{eq:distr3d}).
Linearizing equations (\ref{eq:vmd1}) and (\ref{eq:vmd2}) around VMD ``thermal equilibrium'' distribution function (\ref{eq:distr3d}) one obtains 3D  VMD dispersion relation:
\begin{equation}
\begin{split}
    2Nk^2=\sum_{i=1}^N \left( Z'(\zeta_i)-\tan^2(\theta) \frac{Z(\zeta_i)}{\zeta_i} \right), \\
    \zeta_i=\frac{-|{\bf u}_{i\bot}|\sin(\theta)cos(\varphi_i-\varphi) + \omega/k}{\sqrt{2}\cos(\theta)},
    \end{split}
    \label{eq:vmd3DDisp}
\end{equation}
where $\theta$ and $\varphi$ are polar and azimuthal angle of wave-vector ${\bf k}$ respectively and $\varphi_i=2\pi i/N$ is azimuthal angle of $i$th flow in transverse plane, as depicted in Figure \ref{fig:3Dvectors}.

Similar to $2D$ case, this dispersion relation coincide with the Vlasov equation dispersion relation  in two limiting cases: when $\theta\rightarrow0$ and $\theta\rightarrow\pi/2$.
In the case of $\theta\rightarrow0$, VMD dispersion relation coincide with Vlasov dispersion relation for isotropic thermal plasma \eqref{isotropicVlasovdispersion}.
In the case of $\theta\rightarrow\pi/2$, we obtain a generalized two-stream dispersion relation for cold plasma.
\begin{equation}
    N=\sum_{i=1}^N\frac{1}{(\omega-k|{\bf u}_{i\bot}|cos(\varphi_i-\varphi))^2}.
    \label{eq:twostream3D}
\end{equation}

\begin{figure}
          \includegraphics[width=\plotwidth]{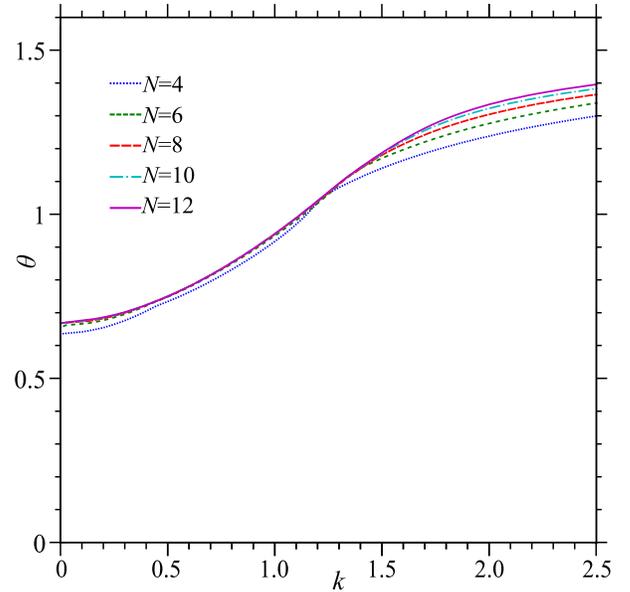}
          \caption{As the number of flows, N, increases, the stability boundary of $3D$ VMD dispersion relation \e{eq:vmd3DDisp} appears to converge.
          Above this boundary, instability persists, but growth rates go to zero as $\sim1/N$.}
          \label{fig:3DstabilityBoundary}
\end{figure}

LW dispersion relation \e{isotropicVlasovdispersion} is qualitatively recovered by \e{eq:vmd3DDisp} for $u\approx\sqrt{2}$ and $\theta\lesssim0.65$. Beyond that angle a variant of the two-stream instability is encountered.
Figure \ref{fig:3DstabilityBoundary} shows the stability region envelopes in angle $\varphi$ (the largest unstable cross-section is chosen over all angles $\varphi$) for
$N=4, 6, 8, 10$ and $12$.  The region below the curves has only stable modes. While the area of the unstable region converges with increase of $N$, the growth rate decreases as $C(k,\theta)/N$, where $C(k,\theta)$ is largest for $\theta\rightarrow \pi/2$ and $C(k,\theta)\rightarrow0$ as we approach the stability boundary or $k=0$ boundary.

\begin{figure}
          \includegraphics[width=3.5in]{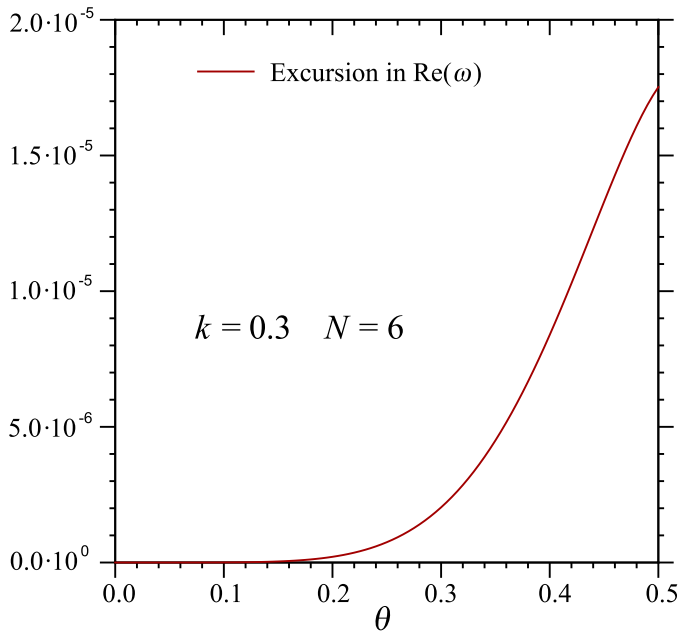}
          \caption{Excursion in real part of LW branch of $3D$ VMD dispersion relation \e{eq:vmd3DDisp} with $N=6$.}
          \label{fig:excursionRE}
\end{figure}

LW dispersion relation \e{isotropicVlasovdispersion} is isotropic for thermal plasma. In general this property is lost for anisotropic equilibria, such as Eq. (\ref{eq:distr3d}),
but it may be approximately regained for small values of the polar angle, $\theta$, for large enough $N$. Based on experience with $2D$ lattice gas models \cite{Frisch}, one expects that six transverse flows,
$N=6$, aligned with the edges of a regular hexagon is sufficient to assure independence of mode properties on the azimuthal angle, $\varphi$, as $\theta\rightarrow0$.

LW branch $\omega_{LW}$ of $3D$ VMD  dispersion relation \e{eq:vmd3DDisp} depends on wavenumber $k$ as well as the polar and azimuthal angles, $\theta$ and $\varphi$.
Figure \ref{fig:excursionRE} shows that the maximum excursion of ${Re}(\omega_{LW})$ as $\varphi$ varies in the interval $0<\varphi \leq\pi/3$,
rapidly goes to zero with $\theta$: at any value of $\theta$, this graph depicts the largest magnitude difference
${Re}(\omega_{LW}(\varphi_1))-{Re}(\omega_{LW}(\varphi_2))$ for $0<\varphi_1,\varphi_2\leq\pi/3$ at
$k=0.3$, normalized to the LW frequency. For equilibrium hexagonal flow geometry, this error varies as $\theta^6$ for small $\theta$, $0<\theta\leq0.2$,
and, since the magnitude of $k$'s projection onto the $xy$ plane, $k_\perp \theta$ for small $\theta$, the error varies as $k_\perp^6$. Similar result can be shown for different number of
transverse flows $N$ arranged uniformly over the azimuthal angle. Excursion of $\text{Re}(\omega_{LW})$ in this case varies as $\theta^N$ for small $\theta$.

Similar to $2D$ case of Section \ref{sec:ACCURACYWITHFLOWNUMBERIN2D}, LW branch of $3D$  VMD  dispersion relation \e{eq:vmd3DDisp} agrees well with the exact
 LW dispersion relation \e{isotropicVlasovdispersion} provided $k\gtrsim 0.3$. The LW branch relative errors between these two dispersion relations are similar to shown in Figures \ref{fig:subfigure1}-\ref{fig:subfigure3}.

\section{CONCLUSION}
\label{sec:Conclusion}

In conclusion, we  investigated the regions of stability of  $2D$ and $3D$ VMD dispersion relations \e{eq:vmd2DNDisp} and \e{eq:vmd3DDisp}, respectively.
We found that with the increase of the number of flows, $N$, these regions quickly converge to the universal curves in $(k,\theta)$ plane. For small $k$ the maximum stable angle is limited to $\theta\lesssim 0.65$.
The dependence of these results on the azimuthal  angle $\varphi$ in $3D$  is  very weak.

We also studied the relative error between Langmuir wave branch of VMD dispersion relations  \e{eq:vmd2DNDisp}, \e{eq:vmd3DDisp} and the exact Langmuir wave dispersion
relation \e{isotropicVlasovdispersion} of the Vlasov equation.
We found that both in $2D$ and $3D$ these errors are small provided $k\gtrsim 0.3$.
E.g., for $k=0.35$ with $N=6$ in $3D$ the relative errors are $\simeq 0.04\%$ for $\theta=0.1$ and $\simeq 2\%$ for $\theta=0.3$.
We also found that in $3D$, the moderate number of flows, $N\ge 6$ already allows us to recover the isotropy of LW dispersion relation for small angles $\theta$ with high precision.

\begin{acknowledgments}
This work was supported by the National Science Foundation
under Grants No. PHY 1004118, and
No. PHY 1004110.
\end{acknowledgments}



\end{document}